\documentclass{PoS}

\title{CPT Symmetry Without Hermiticity}
\ShortTitle{CPT Theorem}
\author{\speaker{Philip D. Mannheim} \\
Department of Physics, University of Connecticut, Storrs, CT 06269, USA\\
\email{philip.mannheim@uconn.edu}}
\date{November 3, 2016}

\abstract{
In the literature the $CPT$ theorem has only been established for Hamiltonians that are Hermitian. Here we extend the $CPT$ theorem to quantum field theories with non-Hermitian Hamiltonians. Our derivation is a quite minimal one as it requires only the time independent evolution of scalar products and  invariance under complex Lorentz transformations. The first of these requirements does not force the Hamiltonian to be Hermitian. Rather, it forces its eigenvalues to either be real or to appear in complex conjugate pairs, forces the eigenvectors of such conjugate pairs  to be conjugates of each other, and forces the Hamiltonian to admit of an antilinear symmetry. The latter requirement then forces this antilinear symmetry to be $CPT$, with Hermiticity of a Hamiltonian thus only being a sufficient condition for $CPT$ symmetry and not a necessary one. $CPT$ symmetry thus has primacy over Hermiticity, and it rather than Hermiticity should be taken as a guiding principle for constructing quantum theories. With conformal gravity being a non-Hermitian theory, our approach allows us to construct a positive, ghost-free norm for the theory, to thereby establish the unitarity of conformal gravity. Since our approach allows for complex energies and decays, our work justifies the use of the $CPT$ theorem in establishing the equality of the lifetimes of unstable particles that are charge conjugates of each other. In the quantum-mechanical limit where charge conjugation is separately conserved, the key results of the $PT$ symmetry program of Bender and collaborators are recovered.}

\FullConference{38th International Conference on High Energy Physics\\
		3-10 August 2016\\
		Chicago, USA}

\begin{document}

\section{Antilinearity and the reality of eigenvalues}

Starting in 1998 with the work of Bender and collaborators (see e.g.  \cite{Bender2007}), it was established that the eigenvalues of the non-Hermitian Hamiltonian $H=p^2+ix^3$ were all real. This reality was traced to the existence of an underlying antilinear $PT$ symmetry that $H$ possessed, a symmetry under which $p\rightarrow p$, $x\rightarrow -x$, $i\rightarrow -i$. To see the relevance of antilinear symmetry, consider  
\begin{eqnarray}
i\partial_ t|\psi(t)\rangle=H|\psi(t)\rangle=E|\psi(t)\rangle.
\label{H1b}
\end{eqnarray}
If we replace the parameter $t$ by $-t$ and multiply by some general antilinear operator $A$, we obtain
\begin{eqnarray}
i\partial_tA|\psi(-t)\rangle=AHA^{-1}A|\psi(-t)\rangle=E^*A|\psi(-t)\rangle.
\label{H2b}
\end{eqnarray}
If $H$ has an antilinear symmetry so that $AHA^{-1}=H$, then, as first noted by Wigner in his study of time reversal invariance,  energies can either be real and have eigenfunctions that obey $A|\psi(-t)\rangle=|\psi(t)\rangle$, or can appear in complex conjugate pairs that have conjugate eigenfunctions ($|\psi(t)\rangle \sim \exp(-iEt)$ and $A|\psi(-t)\rangle\sim \exp(-iE^*t)$). The  converse also holds, since if we are given that the energy eigenvalues are real or appear in complex  conjugate pairs, not only would $E$ be an eigenvalue but $E^*$ would be too. Hence, we can set $HA|\psi(-t)\rangle=E^*A|\psi(-t)\rangle$ in (\ref{H2b}), and obtain
\begin{eqnarray}
(AHA^{-1}-H)A|\psi(-t)\rangle=0.
\label{H3b}
\end{eqnarray}
If the eigenstates of $H$ are complete, (\ref{H3b}) must hold for every eigenstate, to yield $AHA^{-1}=H$ as an operator identity, with $H$ thus having an antilinear symmetry. Since all $E$ real is a special case, and since non-Hermitian Hamiltonians can have a real energy spectrum,  we see that \textbf{antilinearity is a necessary condition for the reality of an eigenspectrum, while Hermiticity is only sufficient.}

\section{A Simple Example}

Consider the $PT$ symmetric matrix $M(s)$ ($P=\sigma_1$ and $T$ is complex conjugation $K$)
\begin{eqnarray}
M(s)=\left(\matrix{1+i&s\cr s&1-i\cr}\right).
\label{H4b}
\end{eqnarray}
Even though this $M(s)$ is not Hermitian, its eigenvalues are given by 
$E_{\pm}=1 \pm (s^2-1)^{1/2}$,
and both of these eigenvalues are real if $s$ is greater than one. Moreover, these eigenvalues come in complex conjugate pairs if $s$ is less than one. Finally,  if $s=1$, then after the similarity transformation
\begin{eqnarray}
\left(\matrix{1+i&1\cr 1&1-i\cr}\right)\rightarrow \left(\matrix{1&0\cr i&1\cr}\right)\left(\matrix{1+i&1\cr 1&1-i\cr}\right)\left(\matrix{1&0\cr -i&1\cr}\right)=\left(\matrix{1&1\cr 0&1\cr}\right),
\label{H6b}
\end{eqnarray}
one can readily check that there is only one eigenvector, viz. $\widetilde{(1,0)}$ where tilde denotes transpose, despite there being two solutions to $|M(1)-\lambda I|=0$ (both with $\lambda=1$). $M(1)$ is thus a non-diagonalizable Jordan-block matrix, to thereby  provide an explicit example of a non-Hermitian operator whose eigenvalues are all real.

As well as see the generic pattern of eigenvalues, we also see that by varying parameters we can continue from one realization of antilinear symmetry to another, crossing through, and in fact necessarily crossing  through, the Jordan-block case on the way, as the transition from all real eigenvalues to complex pairs must be singular. For both $s>1$ and $s<1$ $M(s)$ has a complete set of eigenvectors and can be diagonalized, with its diagonal form being Hermitian when $s>1$. When $s>1$ $M(s)$ is thus ``Hermitian in disguise", with the utility of antilinear  symmetry being that without it, it is guaranteed that a Hamiltonian is not Hermitian in disguise.

The Jordan-block situation is a case where the Hamiltonian is manifestly non-diagonalizable and thus manifestly non-Hermitian and yet all eigenvalues are real. While Hermiticity implies reality of eigenvalues, reality of eigenvalues does not imply Hermiticity or even Hermiticity in disguise. The conformal gravity theory with action $I_{\rm W}=-\alpha_g\displaystyle{\int }d^4x (-g)^{1/2}C_{\lambda\mu\nu\kappa} C^{\lambda\mu\nu\kappa}$ where $C^{\lambda\mu\nu\kappa}$ is the Weyl conformal tensor also falls into the Jordan-block category \cite{Bender2008a,Bender2008b,Mannheim2012a}, and possesses no ghost states of negative norm at the quantum level because of it (the appropriate norm for the theory is the left-right norm discussed below, and it is non-negative \cite{Bender2008a,Bender2008b,Mannheim2012a}), to thus provide a fully consistent and unitary quantum theory of gravity without any need for string theory.

\section{Probability Conservation}

Consider a right eigenstate of $H$ in which $H$ acts to the right as $i\partial_t|R(t)\rangle=H|R(t)\rangle$ with solution $|R(t) \rangle=\exp(-iHt)|R(0) \rangle$. The Dirac norm
\begin{eqnarray}
\langle R(t)|R(t) \rangle=\langle R(0)|\exp(iH^{\dagger}t)\exp(-iHt)|R(0) \rangle
\label{H7b}
\end{eqnarray}
is not time independent if $H$ is not Hermitian, and would not describe unitary time evolution. However, this only means that the Dirac norm is not unitary, not that no norm is unitary. Moreover, since $i\partial_t|R(t)\rangle=H|R(t)\rangle$ only involves ket vectors, there is some freedom in choosing bra vectors. So let us introduce a more general scalar product $\langle R(t)|V|R(t) \rangle$ with some as yet to be determined $V$, which we take to be time independent. We find
\begin{eqnarray}
i\partial_t \langle R_j(t)|V|R_i(t)\rangle =\langle R_j(t)|(VH-H^{\dagger}V)|R_i(t)\rangle.
\label{H8b}
\end{eqnarray}
Thus if we set $VH-H^{\dagger}V=0$, then all scalar products are time independent and probability is conserved (and $V$ will indeed be time independent if $H$ is). For the converse, we note if we are given that all $V$ scalar products are time independent, then if the set of all $|R_i(t)\rangle$ is complete we would obtain $VH-H^{\dagger}V=0$ as an operator identity. The condition $VH-H^{\dagger}V=0$ is thus both necessary and sufficient for the time independence of the $V$ scalar products $\langle R(t)|V|R(t) \rangle$. 

Since $V$ obeys $VH-H^{\dagger}V=0$, $V$ depends on the particular Hamiltonian. Thus unlike the Dirac norm, the theory dynamically determines its own norm each time. This is just like general relativity ($g_{\mu\nu}$ metric determined dynamically) vis a vis special relativity (Minkowski $\eta_{\mu\nu}$ metric preassigned). For the matrix $M(s)$ for instance, we can explicitly construct the $V(s)$ operator:
\begin{eqnarray}
V(s)=\frac{1}{(s^2-1)^{1/2}}\left(\matrix{s&-i\cr i&s\cr}\right),\qquad V(s)M(s)V^{-1}(s)=M^{\dagger}(s),
\label{H9b}
\end{eqnarray}
and expressly note that $V(1)$ is singular (Jordan-block case).

If $VH-H^{\dagger}V=0$, we can set $VH|\psi\rangle=EV|\psi\rangle=H^{\dagger}V|\psi \rangle$. Consequently $H$ and $H^{\dagger}$ have the same set of eigenvalues, i.e. for every $E$ there is an $E^*$. Eigenvalues are thus either real or in complex conjugate pairs. Thus as shown above, \textbf{$H$ must have an antilinear symmetry.}

To reinforce the point we note that if  $|R_i(t) \rangle$ is a right-eigenstate of $H$ with energy eigenvalue $E_i=E_i^R+iE_i^I$, in general we can write
\begin{eqnarray}
&&\langle R_j(t)|V|R_i(t) \rangle
=\langle R_j(0)|V|R_i(0) \rangle e^{-i(E_i^R+iE_i^I)t+i(E_j^R-iE_j^I)t}.
\label{H10b}
\end{eqnarray}
Since $V$ has been chosen so that the $\langle R_j(t)|V|R_i(t) \rangle$ scalar products are time independent,  the only allowed non-zero norms are those that obey
\begin{eqnarray}
&&E_i^R=E_j^R,\qquad E_i^I=-E_j^I,
\label{H11b}
\end{eqnarray}
with all other $V$-based scalar products having to obey $\langle R_j(0)|V|R_i(0) \rangle=0$. We recognize (\ref{H11b}) as being precisely none other than the requirement that eigenvalues be real or appear in complex conjugate pairs, with $H$ thus possessing an antilinear symmetry.

Ordinarily in discussing decays one only keeps modes with negative imaginary part $E^I$. However now we keep both decaying and growing modes, with probability being conserved since the only transitions allowed by (\ref{H11b}) are those in which the decaying mode couples to its growing partner, so that as the population of the decaying mode decreases, the population of the growing mode increases accordingly. Also with $U=e^{-iHt}$ obeying $U^{-1}=e^{iHt}=V^{-1}e^{iH^{\dagger}t}V=V^{-1}U^{\dagger}V$ unitarity is generalized to the non-Hermitian case.

With $VH-H^{\dagger}V=0$ and  $i\partial_t |R \rangle =H|R \rangle$, we obtain  $-i\partial_t \langle R|V =\langle R| H^{\dagger}V=\langle R| VH$. We thus identify left-eigenvectors $\langle L|=\langle R|V$,  with $ \langle L(t)|R(t) \rangle$=$\langle L(0)|e^{iHt}e^{-iHt}|R(0) \rangle$=$\langle L(0)|R(0) \rangle$. For a non-Hermitian $H$ we should use the left-right norm, with it being the most general one possible that is  time-independent. \textbf{Probability conservation thus implies antilinearity not Hermiticity.}

\section{Complex Lorentz Invariance}

Is there any particular antilinear symmetry that might be preferred?  We can get real eigenvalues without assuming Hermiticity. We can get probability conservation without assuming Hermiticity. So can we get the $CPT$ theorem without assuming Hermiticity. Yes,  if we impose complex Lorentz invariance. And if we can get the $CPT$ theorem without Hermiticity, then since energies can then come in complex conjugate pairs, we can now justify the application of the $CPT$ theorem to decays and unstable states such as in the $K$ meson sector, since if one assumes Hermiticity as in the standard derivation of $CPT$ theorem, there then are no decays.

Lorentz transformations are of the form $\Lambda=e^{iw^{\mu\nu}M_{\mu\nu}}$ with six angles $w^{\mu\nu}=-w^{\nu\mu}$ and six Lorentz generators $M_{\mu\nu}=-M_{\nu\mu}$ that obey
\begin{eqnarray}
[M_{\mu\nu},M_{\rho\sigma}]&=&i(-\eta_{\mu\rho}M_{\nu\sigma}+\eta_{\nu\rho}M_{\mu\sigma}
-\eta_{\mu\sigma}M_{\rho\nu}+\eta_{\nu\sigma}M_{\rho\mu}).
\label{H12b}
\end{eqnarray}
Under a Lorentz transformation the line element transforms as 
\begin{eqnarray}
x^{\alpha}\eta_{\alpha\beta}x^{\beta}\rightarrow x^{\alpha}\widetilde{\Lambda}\eta_{\alpha\beta}\Lambda x^{\beta},
\label{H13b}
\end{eqnarray}
with $\widetilde{\Lambda}=e^{iw^{\mu\nu}\widetilde{M_{\mu\nu}}}$. Given the Lorentz algebra, one has $e^{iw^{\mu\nu}\widetilde{M_{\mu\nu}}}\eta_{\alpha\beta}=\eta_{\alpha\beta} e^{-iw^{\mu\nu}M_{\mu\nu}}$ (Minkowski metric orthogonal), with the line element thus being invariant. While this analysis familiarly holds for real $w^{\mu\nu}$, since  $w^{\mu\nu}$ plays no explicit role in it, the analysis equally holds if $w^{\mu\nu}$ is \textbf{complex}.

For a general zero spin Lagrangian $L(x)$ where $w^{\mu\nu}M_{\mu\nu}$ acts as $w^{\mu\nu}(x_{\mu}p_{\nu}-x_{\nu}p_{\nu})$, i.e. as $2w^{\mu\nu}x_{\mu}p_{\nu}$, under an infinitesimal Lorentz transformation  the action $I=\displaystyle{\int} d^4x L(x)$ transforms as 
\begin{eqnarray}
\delta I=2w^{\mu\nu}\int d^4x x_{\mu}\partial_{\nu}L(x)=2w^{\mu\nu}\int d^4x \partial_{\nu}[x_{\mu}L(x)],
\label{H14b}
\end{eqnarray}
to thus be a total derivative and thus be left invariant. Since  the change will be a total derivative even  if $w^{\mu\nu}$ is complex, we again have invariance under \textbf{complex} Lorentz transformations.

For spinors $\psi$ we cannot use $\psi^{\dagger}$ since we would then have to use $\Lambda^{\dagger}=e^{-i[w^{\mu\nu}]^*M^{\dagger}_{\mu\nu}}$. So instead we use Majorana spinors and work in the Majorana basis of the Dirac gamma matrices where $C=\gamma^0$ and all four $\gamma^{\mu}$ are pure imaginary, since in that basis  Majorana spinors are self conjugate. In Grassmann space one has a line element $\widetilde{\psi} C\psi$ where $C$ effects $C\gamma^{\mu}C^{-1}=-\widetilde{\gamma^{\mu}}$ and thus effects $CM_{\mu\nu}C^{-1}=-\widetilde{M_{\mu\nu}}$. We thus obtain
\begin{eqnarray}
\widetilde{\psi} C\psi \rightarrow \tilde{\psi} e^{iw^{\mu\nu}\widetilde{M}_{\mu\nu}}Ce^{iw^{\mu\nu}M_{\mu\nu}}\psi
=\tilde{\psi} Ce^{-iw^{\mu\nu}M_{\mu\nu}}e^{iw^{\mu\nu}M_{\mu\nu}}\psi=\widetilde{\psi} C\psi.
\label{H15b}
\end{eqnarray}
So once again we have invariance under {\bf complex} Lorentz transforms and not just under real ones.

For a general Dirac spinor we set $\psi=\psi_1+i\psi_2$  with $\psi_1=\psi_1^{\dagger}$, $\psi_2=\psi_2^{\dagger}$. Then 
under charge conjugation we obtain $\hat{C}\psi_1\hat{C}^{-1}=\psi_1$, $\hat{C}\psi_2\hat{C}^{-1}=-\psi_2$ (the hat denotes quantum operator). Thus we first set $\bar{\psi} \psi=\psi^{\dagger}\gamma^0\psi=(\tilde{\psi}_1-i\tilde{\psi}_2)C(\psi_1+i\psi_2)$ and then apply a Lorentz transformation, to thus obtain invariance under complex Lorentz transformations. In the Majorana basis of the gamma matrices  $\hat{P}$, $\hat{T}$, and $\hat{C}\hat{P}\hat{T}$ implement 
\begin{eqnarray}
&&\hat{P}\psi(\vec{x},t)\hat{P}^{-1}=\gamma^0\psi(-\vec{x},t),\qquad \hat{T}\psi(\vec{x},t)\hat{T}^{-1}=\gamma^1\gamma^2\gamma^3\psi(\vec{x},-t),
\nonumber\\
&&
\hat{C}\hat{P}\hat{T}[\psi_1(x)+i\psi_2(x)]\hat{T}^{-1}\hat{P}^{-1}\hat{C}^{-1} =i\gamma^5 [\psi_1(-x)-i\psi_2(-x)],
\label{H16b}
\end{eqnarray}
the last relation of which will prove central below.

If $\hat{M}_{\mu\nu}=\hat{M}_{\mu\nu}^{\dagger}$ and $w^{\mu\nu}$ is complex, then $\Lambda^{\dagger}=e^{-i[w^{\mu\nu}]^*M_{\mu\nu}}\neq \hat{\Lambda}^{-1}=e^{-iw^{\mu\nu}M_{\mu\nu}}$, and the Dirac norm $\langle R|R \rangle \rightarrow \langle R|\Lambda^{\dagger}\Lambda|R \rangle$ is not invariant under a complex Lorentz transform. Thus need some additional operator that will convert $e^{-i[w^{\mu\nu}]^*\hat{M}_{\mu\nu}}$ into $e^{-iw^{\mu\nu}\hat{M}_{\mu\nu}}$. It would have to be antilinear in order to complex conjugate $[w^{\mu\nu}]^*$. But since it would also conjugate the $i$ factor it would at the same time have to convert $\hat{M}_{\mu\nu}$ into $-\hat{M}_{\mu\nu}$. Because of the factor $i$ in the Lorentz algebra where $[\hat{M},\hat{M}]=i\hat{M}$, the operator that does this is precisely $\hat{C}\hat{P}\hat{T}$, with $\hat{C}\hat{P}\hat{T}\hat{M}_{\mu\nu}[\hat{C}\hat{P}\hat{T}]^{-1}=-\hat{M}_{\mu\nu}$. Consequently $\langle R|\hat{C}\hat{P}\hat{T}|R \rangle$ is invariant under complex Lorentz transformations when $\hat{M}_{\mu\nu}$ is Hermitian, while $\langle R|V\hat{C}\hat{P}\hat{T}|R \rangle$ is invariant when $V\hat{M}_{\mu\nu}=\hat{M}_{\mu\nu}^{\dagger} V$. This then is how one constructs matrix elements in general that are invariant under complex Lorentz transformations, and as we see, \textbf{complex Lorentz invariance is just as natural to physics as real Lorentz invariance.}

\section{Relation of $CPT$ to Complex Lorentz Transformations}

On coordinates $PT$ implements $x^{\mu}\rightarrow -x^{\mu}$, and thus so does $CPT$ since the coordinates are charge conjugation even (i.e. unaffected by charge conjugation).  With a typical boost in the $x_1$-direction implementing $x_1^{\prime}=x_1\cosh\xi +t\sinh \xi $, $t^{\prime}=t\cosh\xi +x_1\sinh \xi $, on setting $\xi=i\pi$ we obtain  
\begin{eqnarray}
 \Lambda^{0}_{\phantom{0}1}(i\pi):&&~x_1\rightarrow -x_1,~ t\rightarrow -t,\qquad 
 \Lambda^{0}_{\phantom{0}2}(i\pi):~x_2\rightarrow -x_2,~t\rightarrow -t,
 \nonumber\\
 \Lambda^{0}_{\phantom{0}3}(i\pi):&&~x_3\rightarrow -x_3,~t\rightarrow -t,\qquad
 \pi\tau=\Lambda^{0}_{\phantom{0}3}(i\pi)\Lambda^{0}_{\phantom{0}2}(i\pi)\Lambda^{0}_{\phantom{0}1}(i\pi):~x^{\mu}\rightarrow -x^{\mu}.
\label{H17b}
\end{eqnarray}
The complex $\pi\tau$ thus implements the linear part of a $PT$ and thus a $CPT$ transformation on the $x^{\mu}$.

With $\Lambda^{0}_{\phantom {0}i}(i\pi)$ implementing $e^{-i\pi\gamma^0\gamma_i/2}=-i\gamma^0\gamma_i$ in the Dirac gamma space,  we can set
\begin{eqnarray}
&&\hat{\pi}\hat{\tau}=\hat{\Lambda}^{0}_{\phantom{0}3}(i\pi)\hat{\Lambda}^{0}_{\phantom{0}2}(i\pi)\hat{\Lambda}^{0}_{\phantom{0}1}(i\pi)=i\gamma^0\gamma^1\gamma^2\gamma^3=\gamma^5,
\nonumber\\
&&\hat{\pi}\hat{\tau}\psi_1(x)\hat{\tau}^{-1}\hat{\pi}^{-1}=\gamma^5 \psi_1(-x),\qquad \hat{\pi}\hat{\tau}\psi_2(x)\hat{\tau}^{-1}\hat{\pi}^{-1}=\gamma^5 \psi_2(-x).
\label{H19b}
\end{eqnarray}
Up to an overall phase, we recognize $\hat{\pi}\hat{\tau}$ as acting as none other than the \textbf{linear} part of a $CPT$ transformation (cf. (\ref{H16b})). \textbf{$CPT$ is thus naturally associated with the complex Lorentz group}.

With the Lagrangian density $L(x)$ having zero spin, $\hat{\pi}\hat{\tau}$ effects $\hat{\pi}\hat{\tau}L(x)\hat{\tau}^{-1}\hat{\pi}^{-1}=L(-x)$ up to a phase. We will show below that the phase is one. Thus, with $K$ denoting complex conjugation, when acting on a zero spin Lagrangian we can identify $\hat{C}\hat{P}\hat{T}=K\hat{\pi}\hat{\tau}$. On applying $K\hat{\pi}\hat{\tau}$ we  obtain
\begin{eqnarray}
\hat{C}\hat{P}\hat{T}\int d^4x L(x)\hat{T}^{-1}\hat{P}^{-1}\hat{C}^{-1}=K\int d^4x L(-x)K=K\int d^4x L(x)K=\int d^4x L^*(x).
\label{H20b}
\end{eqnarray}
Establishing the $CPT$ theorem is thus reduced to showing that $L(x)=L^*(x)$.

\section{$CPT$ Theorem Without Hermiticity}

Using (\ref{H16b}) we find that the $CPT$ phases of the independent fermion bilinears ($\bar{\psi}\psi$, $\bar{\psi}i\gamma^5\psi$, $\bar{\psi}\gamma^{\mu}\psi$, $\bar{\psi}\gamma^{\mu}\gamma^5\psi$, $\bar{\psi}i[\gamma^{\mu},\gamma^{\nu}]\psi$), and accordingly the phases of all bosons, alternate with spin (i.e. respectively $+1$, $+1$, $-1$, $-1$, $+1$). All net zero spin combinations of fermions and bosons are thus $CPT$ even. Also all are real \cite{Mannheim2015}. Since probability conservation requires an antilinear symmetry, it requires $K\int d^4x L(x)K=\int d^4x L(x)$. Thus from (\ref{H20b}) we infer that that $L(x)=L^*(x)$,  that all the numerical coefficients in $L(x)$ are thus real (all zero spin operator combinations already being real), and that $\int d^4x L(x)$ is $CPT$ invariant \cite{Mannheim2015,Mannheim2016}. \textbf{Thus requiring only probability conservation and complex Lorentz invariance enables us to extend the $CPT$ theorem to the non-Hermitian case. Antilinear $CPT$ symmetry thus has primacy over Hermiticity, and it is antilinearity rather than Hermiticity that should be taken as a guiding principle for quantum theory.}

\section{Implication for the $PT$ Symmetry Program}

In the event that $C$ is separately conserved,  $CPT$ invariance reduces to $PT$ invariance, even if the Hamiltonian is not Hermitian. In such cases we recover the non-Hermitian $PT$ program of Bender and collaborators, to thus put it on a quite firm theoretical foundation.

\end{document}